          \documentclass{cmslatex}
\usepackage[paperwidth=7in, paperheight=10in, margin=.875in]{geometry}

\renewcommand{\theequation}{\arabic{section}.\arabic{equation}}

\usepackage{amssymb,amsmath,bm,euscript}
\usepackage{graphicx,color,caption}

\usepackage[colorlinks=true]{hyperref}

\usepackage{algorithm,algpseudocode}
\usepackage{braket}

\newcommand{\bJ}{{\bf J}}
\newcommand{\tr}{\text{tr}}

\newcommand{\bn}{\hat{{\mathbf n}}}
\newcommand{\bv}{{\mathbf v}}

\renewcommand{\theequation}{\thesection.\arabic{equation}}
\newcommand{\qed}{\hphantom{.}\hfill $\Box$\medbreak}

\def\O{\mathcal{O}}

\def\A{\mathcal{A}}

\def\B{\mathcal{B}}
\def\C{\mathcal{C}}

\def\x{{\bf x}}

\def\uu{{\bf u}}
\def\vv{{\bf v}}
\def\0{{\bf 0}}
\def\ketbra#1{|#1\rangle\langle#1|}

\begin{document}
          \title{Regularly Decomposable Tensors and Classical Spin
    States\thanks{This research was supported by the Hong Kong
    Research Grant Council (Grant No.  PolyU 501212, 501913, 531213, 15302114,
    15300715, 15206915, the National Natural Science Foundation of
    China (11401428,61374057), and the Deutsch-Franz\"osische
    Hochschule (Universit\'e franco-allemande), grant number
    CT-45-14-II/2015. Ce travail a b\'en\'efici\'e d'une aide
    Investissements d'Avenir du LabEx PALM (ANR-10-LABX-0039-PALM).}}


\author{Liqun Qi\thanks{Department of Applied
    Mathematics, The Hong Kong Polytechnic University, Hung Hom,
    Kowloon, Hong Kong; (liqun.qi@polyu.edu.hk).}
\and{Guofeng Zhang\thanks{Department of Applied Mathematics, The Hong
    Kong Polytechnic University, Hung Hom, Kowloon, Hong Kong; (magzhang@polyu.edu.hk).}}
  \and{Daniel Braun\thanks{Institut f\"ur theoretische Physik,
    Universit\"at T\"ubingen, 72076 T\"ubingen, Germany; (daniel.braun@uni-tuebingen.de).}}
\and{Fabian Bohnet-Waldraff\thanks{Institut f\"ur theoretische Physik, Universit\"at
    T\"ubingen, 72076 T\"ubingen, Germany; and LPTMS, CNRS,
    Univ.~Paris-Sud, Universit\'e
    Paris-Saclay, 91405 Orsay, France; (fabian.bohnet-waldraff@uni-tuebingen.de).}}
  \and{Olivier Giraud\thanks{LPTMS, CNRS, Univ.~Paris-Sud, Universit\'e
    Paris-Saclay, 91405 Orsay, France;
      (olivier.giraud@lptms.u-psud.fr).}}
}





         \pagestyle{myheadings} \markboth{Regularly Decomposable Tensors and Classical Spin
    States}{Liqun Qi, Guofeng Zhang, Daniel Braun, Fabian Bohnet-Waldraff, Olivier Giraud} \maketitle

          \begin{abstract}
 A spin-$j$ state can be represented by a
symmetric tensor of order $N=2j$ and dimension $4$.    Here, $j$ can
be a positive integer, which corresponds to a boson; $j$ can also be a
positive half-integer, which corresponds to a fermion.    In this paper,
we introduce regularly decomposable tensors and show that a spin-$j$
state is classical if and only if its representing tensor  is a regularly
decomposable tensor.    In the even-order case,
a regularly decomposable tensor is a completely decomposable tensor but
not vice versa; a completely decomposable tensors is a sum-of-squares
(SOS) tensor but not vice versa; an SOS tensor is a positive
semi-definite (PSD) tensor but not vice versa.   In the odd-order case, the
first row tensor of a regularly decomposable tensor is regularly
decomposable and its other row tensors are induced by the regular
decomposition of its first row tensor.
We also show that complete decomposability and regular
decomposability are invariant under orthogonal transformations, and that the
completely decomposable tensor cone and the regularly decomposable
tensor cone are closed convex cones.  Furthermore, in the even-order
case, the completely decomposable tensor cone and the PSD tensor cone are
dual to each other.   The Hadamard product of two completely
decomposable tensors is still a completely decomposable tensor.
Since one may apply the positive semi-definite programming algorithm
to detect whether a symmetric tensor is an SOS tensor or not, this gives a
checkable  necessary condition for classicality of a spin-$j$ state.
Further research issues on regularly decomposable tensors are also
raised.
          \end{abstract}
\begin{keywords}
positive semi-definite tensors, sum-of-squares tensors, quantum entanglement, spin states, bosons, fermions, classicality
\end{keywords}

 \begin{AMS}
 15A18, 15A69, 15B48
\end{AMS}

\section{Introduction}


A geometrical picture of quantum states often helps getting some
insight on underlying physical properties. For arbitrary pure spin
states, such a geometrical representation was developed by Ettore
Majorana \cite{Majorana}: a spin-$j$ state is visualized as $N=2j$
points on the unit sphere $S^2$, called in this context the Bloch sphere. The advantage of such a picture is a direct interpretation of certain unitary operations: namely, if a quantum spin-$j$ state is mapped to another one by a unitary operation that correspond to a $(2j+1)$-dimensional representation of a spatial rotation, its Majorana points are mapped to points obtained by that spatial rotation.
Recently a tensor representation of an arbitrary mixed or pure spin-$j$ state was proposed that generalizes this picture \cite{GBBBM}.
It consists of a real symmetric tensor of
order $N=2j$ and dimension $4$.   A spin-$j$ state corresponds to a boson if $j$ is a positive integer, and corresponds to a fermion if $j$ is a positive half-integer.
 Thus, a boson corresponds to an even-order four dimensional tensor, while a fermion corresponds to an odd
order four dimensional tensor.

The geometrical picture is particularly useful when it comes to
studying classicality properties of spin states. In quantum optics,
coherent states are quantum states that behave the most classically,
in that they minimize the uncertainty relation between position and
momentum. Coherent states can also be defined in the context of
spins. Statistical mixtures of coherent states can thus be considered
the ``least
 quantum'' states. The set of classical spin states
was introduced in \cite{Giraud08} as the convex hull of the set of coherent spin states.  It can be interpreted (see e.g.~\cite{BBG}) as the set of fully separable states in the symmetric sector of the tensor product of $2j$ spins-$1/2$.
The above geometric picture easily allows one
to characterize coherent spin states: a coherent spin-$j$ state can be represented by $N=2j$ points located at the same position on the Bloch sphere. The characterization of classical states is less easy to obtain, but the tensorial picture helps to get some results on this issue. For instance, in \cite{BBG} it was shown that when $j$ is an
integer, i.e., $N$ is an even number, a classical spin-$j$ state is
such that its representing tensor  is positive semi-definite
(PSD) in the sense of \cite{Qi} { (see Section \ref{sec:defs}).

Positive semi-definiteness of the tensor representation is a necessary
and sufficient condition { of classicality} 
in the case $j=1$ \cite{qspin1}. It is only a
necessary condition for classicality of a spin-$j$ state, and only if
$j$ is a positive integer, as pointed out in \cite{BBG}. A natural
question is therefore whether it is possible to formulate a necessary
and sufficient condition for classicality of a spin-$j$ state in terms
of its tensor representation, first in the case where $j$ is a
positive integer, i.e., the boson case, and then in the case where $j$
is a half-integer, i.e., the fermion case. 
The aim of this paper is to introduce tools in order to reformulate these two questions from a mathematical perspective.

The PSD condition can be expressed in terms of tensor eigenvalues. A tensor is PSD if and only if its smallest
H-eigenvalue or Z-eigenvalue is nonnegative \cite{Qi}.  This links
classicality of a spin-$j$ state (with $j$ as a positive integer) with
the smallest tensor eigenvalue of its representing tensor. 
This result
echoes the result of \cite{HQZ}, which stated that
the geometric measure of entanglement of a pure state is equal to the
largest tensor eigenvalue.  Note that tensor eigenvalues have found
applications in different areas of physics \cite{GV, Ka, Ra, Vi}. To go beyond the PSD condition for classicality, we have to consider stronger properties.}
A property stronger than positive semi-definiteness is the sum-of-squares (SOS) property.  SOS tensors were introduced in \cite{HLQ, LQY}.  According to the Hilbert theory \cite{Hi}, an SOS tensor is a PSD tensor but not vice versa.    Both PSD and SOS tensors have been studied intensively in recent years.   Some references on PSD and SOS tensors include \cite{CLQ, CQ, LWZZL, Qi15, QS, QXX, SQ, ZQZ}. One can show (see below) that when $j$ is an integer, if a spin-$j$ state is classical, then its representing tensor is an SOS tensor in the sense of \cite{CLQ, HLQ, LQY}. But this is still a necessary condition.
A property stronger than the SOS property is complete decomposability.
Completely decomposable tensors were introduced and studied in
\cite{LQX, WLQX}.   An even-order completely decomposable 
tensor is an SOS tensor but not vice versa \cite{LQX, WLQX}.   Again, when $j$ is an integer, if a spin-$j$ state is classical, then its representing tensor is a completely decomposable tensor, and this is still a necessary condition.

In this paper, we introduce regularly decomposable tensors.   A
regularly decomposable tensor is a completely decomposable tensor but
not vice versa.   Furthermore, we define regularly decomposable
tensors also in the odd-order case.   In the odd-order case, the first row
tensor of a regularly decomposable tensor is regularly decomposable
and its other row tensors are induced by the regular decomposition of
its first row tensor.   We show that in both the odd-order (fermion) and
even-order (boson) cases a spin-$j$ state is classical if and only if its
representing tensor 
is a regularly decomposable tensor.    Thus, it is important to study
properties of regularly decomposable tensors and completely
decomposable tensors, as well as some further properties of PSD
tensors and SOS tensors.

The remaining part of this paper is organized as follows.    In Section \ref{sec:defs}, we review the definitions of PSD, SOS and completely
decomposable tensors, and define regularly decomposable tensors.   In
Section 3, we show that in both the odd-order (fermion) and even-order (boson)
cases a spin-$j$ state is classical if and only if its representing
tensor
is a regularly decomposable tensor.    Some properties of
completely decomposable tensors and regularly decomposable tensors and
their implications in physics are studied in Section 4.   Some further
research issues on regularly decomposable tensors are raised in Section
5.

\section{PSD, SOS, Completely Decomposable and Regularly Decomposable Tensors}
\label{sec:defs}
In this paper, for a vector $\x \in \Re^{n+1}$, we denote it as $\x =
(x_0, x_1, \ldots, x_n)^\top$.    Later, in physical applications,
we will have $n=3$.   Here, we assume that $n \ge 2$.   Denote the
zero vector in $\Re^{n+1}$ by $\0$.

Let $\A = (a_{i_1\ldots i_m})$ be an $m$th order $(n+1)$-dimensional
real tensor.  We say that $\A$ is a symmetric tensor if the entries
$a_{i_1\ldots i_m}$ are invariant under permutation of their indices.
Denote $T_{m, n+1}$ as the set of all $m$th order $(n+1)$-dimensional
real tensors,  and $S_{m, n+1}$ as the set of all $m$th order
$(n+1)$-dimensional real symmetric tensors.  Then $T_{m, n+1}$ is a
linear space, and $S_{m, n+1}$ is a linear subspace of $T_{m, n+1}$.
Denote the zero tensor in $S_{m, n+1}$ by $\O$.

 Let $\A = (a_{i_1\ldots i_m}) \in T_{m, n+1}$ and $\B = (b_{i_1\ldots
   i_p}) \in T_{p, n+1}$.  The outer product of $\A$ and $\B$, denoted
 as $\C = \A \otimes \B$, is a real tensor in $T_{m+p, n+1}$, defined
 by $\C = (a_{i_1\ldots i_m}b_{i_{m+1}\ldots i_{m+p}})$.   We also
 denote $\A^{\otimes 2} = \A \otimes \A$, $\A^{\otimes (k+1)} =
 \A^{\otimes k} \otimes \A$ for $k \ge 2$. 
{ A {\em symmetric rank-one} tensor is defined
 as a symmetric tensor in $S_{m, n+1}$ of the form
 $\alpha \x^{\otimes m}$, where $\alpha \in \Re$ and $\x \in \Re^{n+1}$}.

 Let $\A = (a_{i_1\ldots i_m})$ and $\B = (b_{i_1\ldots i_m})$ in
 $S_{m, n+1}$.   The inner product of $\A$ and $\B$, denoted as $\A
 \bullet \B$, is
 a scalar, defined by
 $$\A \bullet \B = \sum_{i_1, \ldots, i_m=0}^n a_{i_1\ldots i_m}b_{i_1\ldots i_m}.$$

Let $\A = (a_{i_1\ldots i_m}) \in S_{m, n+1}$ and $\x \in \Re^{n+1}$.   Then we have
$$\A \bullet \x^{\otimes m} \equiv \sum_{i_1, \ldots, i_m=0}^n a_{i_1\ldots i_m}x_{i_1}\ldots x_{i_m}.$$
If for any $\x \in \Re^{n+1}$, we have $\A \bullet \x^{\otimes m} \ge
0$, then we say that $\A$ is a {\bf positive semi-definite (PSD)}
tensor.    If for any $\x \in \Re^{n+1}, \x \not = \0$, we have $\A
\bullet \x^{\otimes m} > 0$, then we say that $\A$ is a {\bf positive
  definite (PD)} tensor.   Clearly, if $m$ is odd, then the only PSD
tensor is the zero tensor, and there is no PD tensor.    Thus, we only
discuss even-order PSD and PD tensors.

Suppose that $m=2l$ is even.   Let $\A \in S_{m, n+1}$.   If there are
symmetric tensors $\A^{(1)}, \ldots, \A^{(r)} \in S_{l, n+1}$ such
that for all $\x \in \Re^{n+1}$,
$$\A \bullet \x^{\otimes m}= \sum_{k=1}^r \left(\A^{(k)} \bullet \x^{\otimes l} \right)^2,$$
then $\A$ is called a {\bf sum-of-squares (SOS)} tensor.   Then, for
any $\x \in \Re^{n+1}$, we have $\A \bullet \x^{\otimes m} \ge 0$.
Thus, an SOS tensor is always a PSD tensor, but not vice versa.   By
the Hilbert theory \cite{Hi}, only in the following three cases: 1)
$m=2$, 2) $n=1$, 3) $m=4$ and $n=2$, a PSD tensor is always an SOS
tensor; otherwise, there are always PSD tensors which are not SOS
tensors.   David Hilbert \cite{Hi} stated this in the language of
polynomials.   But the meanings are the same.

Let $\A \in S_{m, n+1}$. { Here, $m$ can be either even or odd.}
If there are vectors $\uu^{(1)}, \ldots,
\uu^{(r)} \in \Re^{n+1}$ such that
\begin{equation} \label{e1}
\A = \sum_{k=1}^r \left(\uu^{(k)}\right)^{\otimes m},
\end{equation}
then we say that $\A$ is a {\bf completely decomposable tensor}.   If all the vectors $\uu^{(1)}, \ldots,
\uu^{(r)} \in \Re^{n+1}$ are nonnegative vectors, then $\A$ is called
a {\bf completely positive} tensor \cite{LQ, QXX}.
Actually, all odd-order symmetric tensors are completely decomposable
tensors \cite{LQX}.   Thus, the concept of completely decomposable tensors
is not useful for odd order.   However, if $m=2l$ is even, and $\A$ is a
completely decomposable tensor as defined by (\ref{e1}), then by
letting $\A^{(k)} = \left(\uu^{(k)}\right)^{\otimes l}$, we see that
$\A$ is an SOS tensor.  On the other hand, by the examples given in
\cite{LQX, WLQX}, an SOS tensor may not be a completely decomposable
tensor.

In order to define regularly decomposable tensors, we still need two
more concepts: regular vectors and row-tensors.
\begin{definition}
Let $\x = (x_0, x_1, \ldots,
x_n)^\top \in \Re^{n+1}$.   We say that $\x$ is a {\bf regular vector}
if $x_0 \not = 0$ and $x_0^2 = x_1^2 + \ldots + x_n^2$.
\end{definition}
\begin{definition}
For any $\A = (a_{i_1\ldots i_m}) \in S_{m, n+1}$, define its $i$th
{\bf row tensor} $\A_i$ as a symmetric tensor in $S_{m-1, n+1}$, by
$\A_i = (a_{ii_2\ldots i_m})$, for $i = 0, \ldots, n$.
\end{definition}

We can then define regularly decomposable tensors as follows:
{
\begin{definition}
{\em (i.)} Let the order $m=2l$ be even and $\A \in S_{m, n+1}$.  If $\A$ is a completely
decomposable tensor defined by
(\ref{e1}), where $\uu^{(1)}, \ldots, \uu^{(r)}$ are regular vectors, then we say that $\A$ is a {\bf regularly decomposable tensor} of even order. \\
{(\em ii.)} Let the order $m=2l+1$ be odd and $\A \in S_{m, n+1}$.    If $\A_0
\in S_{2l, n+1}$ 
is a regularly decomposable tensor with the regular
decomposition
\begin{equation} \label{e3}
\A_0 = \sum_{k=1}^r \left(\uu^{(k)}\right)^{\otimes 2l},
\end{equation}
where $\uu^{(k)} = \left(u^{(k)}_0, \ldots, u^{(k)}_n\right)^\top$, $k
= 1, \ldots, r$, are regular vectors, and the
other row tensors of $\A$ are induced by this regular decomposition,
\begin{equation} \label{e4}
\A_i = \sum_{k=1}^r {u^{(k)}_i \over u^{(k)}_0}\left(\uu^{(k)}\right)^{\otimes 2l},
\end{equation}
for $i = 1, \ldots, n$, then we say that $\A$ is a {\bf regularly
  decomposable tensor} of odd order.
\end{definition}
 Clearly an even-order regularly decomposable tensor is a completely
 decomposable tensor but not vice versa.


\begin{theorem} \label{t1}
A regularly decomposable tensor $\A = (a_{i_1\ldots i_m}) \in S_{m, n+1}$ can be written as
\begin{equation} \label{e5}
\A = \sum_{k=1}^r \alpha_k \left(\vv^{(k)}\right)^{\otimes m},
\end{equation}
where $\alpha_k > 0$ and $\vv^{(k)} = \left(1, v^{(k)}_1, \ldots, v^{(k)}_n\right)^\top$,
\begin{equation} \label{e6}
\sum_{i=1}^n \left(v^{(k)}_i\right)^2 =1,
\end{equation}
for $k = 1, \ldots, r$.   
Furthermore, we have\begin{equation} \label{e8}
a_{00i_3\ldots i_m} = \sum_{i=1}^n a_{ii\,i_3 \ldots i_m}
\end{equation}
 for $m\ge 2$ and all $i_3, \ldots, i_m=0, 1, \ldots, n$.
\end{theorem}

{\bf Proof.} Suppose that $m$ is even, and $\A$ is defined by
(\ref{e1}), where $\uu^{(1)}, \ldots, \uu^{(r)}$ are regular vectors.  Let
\begin{equation}
  \label{eq:vv}
\vv^{(k)} = {\uu^{(k)} \over u^{(k)}_0},
\end{equation}
for $k = 1, \ldots, r$.   Then we see that $\A$ can be expressed by
(\ref{e5}), where $\alpha_k = \left(u^{(k)}_0\right)^m > 0$ and $\vv^{(k)} = \left(1, v^{(k)}_1,
  \ldots, v^{(k)}_n\right)^\top$ satisfy (\ref{e6}) for $k = 1,
\ldots, r$.
Suppose that $m=2l+1$ is odd, and $\A_0$ is defined by (\ref{e3}),
where $\uu^{(1)}, \ldots, \uu^{(r)}$ are regular vectors and the other row tensors of $\A$ are defined by
(\ref{e4}).   Then we see that $\A$ can also be expressed by
(\ref{e5}), where $\alpha_k = \left(u^{(k)}_0\right)^{2l} > 0$, 
$\vv^{(k)} = \left(1, v^{(k)}_1,
  \ldots, v^{(k)}_n\right)^\top$, still defined by
\eqref{eq:vv},
 satisfy (\ref{e6}) for $k = 1,
\ldots, r$. 
By these, we see that (\ref{e8}) is satisfied.   
\qed

Suppose that $\A = (a_{i_1\ldots i_m}) \in S_{m, n+1}$ satisfies (\ref{e8}).   Then we call $\A$ a {\bf regular
  symmetric tensor}. If moreover $a_{00\ldots 0} =1$ we call $\A$ a {\bf regular normalized
  symmetric tensor}. In the next section we will see that an important
research issue is to determine whether
a given regular symmetric tensor is a regularly decomposable tensor or
not.}

\section{Regularly Decomposable Tensors and Classicality of Spin States}
Several definitions of classicality of a quantum state exist in the
literature, based e.g.~on the positivity of the Wigner function, or
the absence of entanglement in the case of multi-partite
systems \cite{Richter02,Kenfack04,Cormick06,HQZ,HQSZ}.  In
\cite{Giraud08} a
suitable definition of classicality of spin states was
introduced. Firstly, pure {classical}
spin states are defined as
angular-momentum coherent states, also called ``SU(2)-coherent
states'',
and in the following { also} simply ``coherent states''. Their
properties are
well-known from work in quantum optics
\cite{Arecchi72,Agarwal81} and quantum-chaos \cite{Haake00}.
For being self-contained, we briefly
review them here.\\

{ SU(2)-coherent states}
 can be labeled by a complex label $\alpha$, related by
stereographic projection to polar and azimuthal angles $\theta$ and
$\phi$, $\alpha=\tan(\theta/2)e^{\imath\phi}$ with $\theta \in [0,\pi]$ and $\phi \in [0,2\pi[$. Let $\bJ  \equiv
  (J_x,J_y,J_z)$ denote the  angular momentum vector, and $\ket{j,m}$
  the joint-eigenbasis { states}
of the angular momentum component $J_z$ and the total angular momentum
$\bJ^2\equiv J_x^2+J_y^2+J_z^2$, with
$J_z\ket{j,m}=m\ket{j,m}$, $\bJ^2\ket{j,m}=j(j+1)\ket{j,m}$.  The
components $J_x$ and $J_y$ are related to the ladder operators $J_\pm$
by $J_\pm=J_x\pm \imath J_y$ and $J_\pm \ket{j,m}=\sqrt{j(j+1)-m(m\pm
  1)}\ket{j,m\pm 1}$, {where $\imath =
\sqrt{-1}$ is the imaginary unit.}
 The coherent
states can be written as
\begin{equation}
\label{spincoherent}
\ket{\alpha} =\!\!\! \sum_{m=-j}^j \sqrt{\binom{2j}{j+m}} \left(\cos\frac{\theta}{2}\right)^{j+m}\left(\sin\frac{\theta}{2}e^{\imath\phi}\right)^{j-m}\!\!\!\!\!\! \ket{j,m}.
\end{equation}
 For $\theta=0$ or
$\theta=\pi$, $\ket{\alpha}=\ket{j,j}$ or $\ket{j,-j}$ respectively,
i.e.~the angular momentum states with largest or smallest
$J_z$-component are always coherent states. Geometrically, a coherent state $|\alpha\rangle$ with $\alpha = \tan(\theta/2)e^{\imath\phi}$  is associated to a direction  $\bn =(\sin\theta\,\cos\phi,\sin\theta\,\sin\phi,\cos\theta)$ on the Bloch sphere.   Coherent states have the
important property that the quantum uncertainty of the rescaled angular
momentum vector $\bJ/j$ of a spin-$j$ is minimal for all pure quantum states,
$(\braket{\alpha|\bJ^2|\alpha}-\braket{\alpha|\bJ|\alpha}^2)/j^2=1/j$.  The uncertainty vanishes in  the classical limit of a large
spin, $j\to\infty$.  The coherent states come as closely as
possible to the ideal of a classical phase space point, i.e.~represent
as best as allowed by the laws of quantum mechanics an angular
momentum pointing in a precise direction,
\begin{equation}
  \label{eq:Ja}
\braket{\alpha|\bJ|\alpha}=j(\sin\theta\,\cos\phi,\sin\theta\,\sin\phi,\cos\theta) = j\bn\,.
\end{equation}
\\
Another important feature of coherent states is that they remain
coherent under unitary transformations of the form $U=e^{-\imath\gamma
  \bn\cdot\bJ}$. Such unitary transformations arise from the dynamics
of the angular momentum in a magnetic field (assuming that the angular
momentum is associated with a magnetic moment).  Classically, the spin
precesses around the axis given by the magnetic field, and this is
reproduced by the behavior of the coherent state. One can see this
most easily for $\bn=\hat{e}_z  = (0,0,1)$, i.e.~a magnetic field in the
$z$-direction, in which case $U=e^{-\imath\gamma
  J_z}$ can be immediately applied to the basis states
$\ket{j,m}$ and gives rise to additional phase factors $e^{-\imath
  \gamma m
  }$, i.e.~$\phi\mapsto\phi+\gamma$, and correspondingly the
expectation value $\braket{\alpha|\bJ|\alpha}$ is rotated by the angle
$\gamma$ about the $z-$axis. In general, the mapping
$\ket{\alpha}\mapsto \ket{\tilde{\alpha}}=e^{-\imath
  \gamma \bn\cdot \bJ}\ket{\alpha}$ leads to an expectation value
$\braket{\tilde{\alpha}|\bJ|\tilde{\alpha}}=R(\bn,\gamma)\braket{\alpha|\bJ|\alpha}$,
where $R(\bn,\gamma)$ is a $3\times 3$ orthonormal matrix representing
rotation about the axis $\bn$ with a rotation
angle $\gamma$. Due to Eq.~\eqref{eq:Ja}, it is clear that all coherent states can
be obtained by an appropriate
unitary transformation of the form $U=e^{-\imath  \gamma \bn\cdot\bJ}$
acting on the state $\ket{j,j}$ associated with the direction
$\hat{e}_z$. \\

{
The quantum state of any physical system with finite dimensional
Hilbert space can be represented by a density operator (also called
density matrix)
$\rho$, a positive semi-definite hermitian operator with $\tr\rho=1$.
{ If $\lambda_i$ and $\ket{\psi_i}$ are respectively the eigenvalues and eigenvectors of $\rho$, one has the eigendecomposition $\rho=\sum_i\lambda_i\ket{\psi_i}\bra{\psi_i}$. The density matrix $\rho$ can therefore be interpreted as representing a quantum state which is in some pure state $\ket{\psi_i}$ with probability $\lambda_i$. The condition $\tr\rho=1$ ensures that the probabilities are normalized to 1; it is however possible to work with unnormalized density matrices by relaxing the constraint on $\tr\rho$. In the present paper we will follow that option. As most equations we consider are linear in $\rho$, this just means that we may forget about an overall normalization constant.}

The density operator of an arbitrary spin-$j$ quantum state can be
written in the form of a diagonal
representation,
\begin{equation}
  \label{eq:P}
  \rho=\int_{S^2} d\alpha P(\alpha)\ketbra{\alpha}\,,
\end{equation}
where $P(\alpha)$ is known as the (Glauber-Sudarshan) $P-$function
\cite{Agarwal81}, and $d\alpha=\sin\theta d\theta d\phi $ is the
integration measure over the unit sphere $S^2$ in three dimensions.
Classically mixing states, i.e.~drawing randomly pure states according
to a classical probability distribution, should not increase the
non-classicality
of a state.  Hence, a spin-state is called classical, if and only if
a decomposition of $\rho$ in the form of Eq.~\eqref{eq:P} exists with
$P(\alpha)\ge 0$, in which case $P(\alpha)$ can be interpreted as
classical probability density of finding the pure SU(2)-coherent state
$\ket{\alpha}$ in the mixture. Since by definition
classical states form a convex set,
Caratheodory's theorem implies immediately that a classical state can
be written as a finite convex sum of projectors onto coherent
states,
{
\begin{equation}
  \label{eq:Pw}
  \rho=\sum_{i=1}^{(N+1)^2+1}w_i\ketbra{\alpha_i}\,,
\end{equation}
where $w_i\ge 0$}. Eq.~\eqref{eq:Pw} is the general definition of a
classical spin state adopted in \cite{Giraud08}, and we will base the
rest of the paper on it.\\

A single spin-1/2 is equivalent to a qubit, i.e.~a quantum-mechanical
two state system.  The two states ``spin-up''  and ``spin-down'',
namely $\ket{\frac{1}{2},\frac{1}{2}}$ and
$\ket{\frac{1}{2},-\frac{1}{2}}$ in the above $\ket{j,m}$ notation, are
also called ``computational-basis''.  Denoted as $\ket{0}$ and
$\ket{1}$ in quantum-information theory, they are represented as
column-vectors $(1,0)^T$ and $(0,1)^T$.  In this basis, the density
operator can be represented by a $2\times 2 $ complex
hermitian matrix with $\tr\rho=1$ that can be expanded over the Pauli-matrix basis,
$$\sigma_0 = \left({\begin{array} {cc} 1
      & 0 \\ 0 & 1 \end{array}}\right), \ \sigma_1 =
\left({\begin{array} {cc} 0 & 1 \\ 1 & 0 \end{array}}\right),
 \ \sigma_2 = \left({\begin{array} {cc} 0 & -\imath \\ \imath &
       0 \end{array}}\right) \ {\rm and}\ \sigma_3 =
 \left({\begin{array} {cc} 1 & 0 \\ 0 & -1 \end{array}}\right),$$
\begin{equation}
  \label{eq:bloch}
  \rho=\frac{1}{2}\sum_{i=0}^3\sigma_i a_i\,.
\end{equation}
The four components $a_i$, $i\in\{0,1,2,3\}$ form an order-1 tensor
$\mathcal A$ of dimension 4. The Pauli matrices $
(\sigma_1,\sigma_2,\sigma_3) \equiv\bm\sigma$ are matrix representations of the
{ components of the operator} $2\bJ$
in the  ``spin-up''  and
``spin-down'' computational-basis.  We have $\tr\rho=a_0$. { The vector }
$\bv\equiv (a_1,a_2,a_3)^T\in \Re^3$ is the so-called Bloch
vector. It  satisfies
{ $||\bv||_2\le a_0$} in order to guarantee the positivity of $\rho$.  In
particular, { $||\bv||_2=a_0$} signals pure states (i.e.~rank-1 states), and
{ $||\bv||_2<a_0$} mixed states (rank-2 states). Due to the
orthonormality of the Pauli-matrix basis, $\bv$ can be obtained from a
given state as $\bv=\tr\rho\bm\sigma$. In particular, for a SU(2)-coherent
state $\ket{\alpha}$, one finds
$\bv=\braket{\alpha|2\bJ|\alpha}=(\sin\theta\,\cos\phi,\sin\theta\,\sin\phi,\cos\theta)$, as evidenced by Eq.~(\ref{eq:Ja}).
The   Bloch picture is particularly useful for visualizing unitary
operations: Due to the rotation properties of the coherent states under a unitary
operation, if $\tilde{\rho}=U\rho U^\dagger$, the corresponding Bloch
vector $\tilde{\bv}$ of
$\tilde{\rho}$ is obtained by rotation of
the original Bloch vector, namely $\tilde{\bv}=R(\bn,\gamma)\bv$.  
As the zero-component of tensor $\mathcal{A}$  has to
remain unchanged { due to the conservation of the trace under
  unitary operations,
$\tilde{a}_0=a_0$},
the transformation of $\mathcal{A}$ reads
$\tilde{a}_i=\mathcal{R}_{ij}a_j$ with
\begin{equation}
  \mathcal{R}_{00}=1,\,\,\,
\mathcal{R}_{0i}=\mathcal{R}_{i0}=0 \mbox{ ($i\in\{1,2,3\}$) and }
\mathcal{R}_{ij}=R(\bn,\gamma)_{ij}\mbox{
  ($i,j\in\{1,2,3\}$). } \label{Rs} 
\end{equation}

In \cite{GBBBM} the Bloch-vector of a spin-1/2 was generalized to a
Bloch-tensor of a spin-$j$.
 A spin-$j$ can be composed from $N=2j$
 spins-1/2. The total spin
is then the sum of the $N$ spins-1/2, i.e.~$\bJ=\sum_{i=1}^N \bm\sigma^{(i)}/2$.
 In
 general, combining two spins $j_1$ and $j_2$ gives rise to total
 spins $j$ ranging from $|j_1-j_2|$ to $j_1+j_2$.  A spin
 $j=N/2$ is hence  the maximum total spin
 achievable with $N$ spins-1/2. All basis states $\ket{j,m}$ can be
 created by acting with the ladder operator $J_-$
on
the state $\ket{j,j}$, which in turn is the state
$\ket{\frac{1}{2} ,\frac{1}{2}}^{\otimes N}$ of all spins-up
in
the full Hilbert space of $N$ spins-$1/2$.  Since
both $\ket{j,j}$ and $J_-$ are fully symmetric under the exchange of
all spins-1/2, all $\ket{j,m}$ states lie in the fully symmetric
subspace $\mathcal{H}_S$
of the total Hilbert-space $\mathcal{H}=\mathbb{C}^{2^{N}}$. A
projector
$\mathcal{P}_S$ onto that subspace can be obtained as
\begin{equation}
\mathcal{P}_{S}\equiv \sum_{k=0}^{N}\left\vert D_{N}^{(k)}\right\rangle
\left\langle D_{N}^{(k)}\right\vert\,,   \label{eq:Ps}
\end{equation}%
where the so-called Dicke states $\ket{D_{N}^{(k)}}$ are defined as
\begin{equation*}
\left\vert D_{N}^{(k)}\right\rangle =\mathcal{N}\sum_{\pi }\vert
\underbrace{0\ldots 0}_{k}\underbrace{1\ldots 1}_{N-k}\rangle ,\ \ \
k=0,\ldots N,
\end{equation*}%
$\mathcal{N}$ is a normalization constant, and the sum is over all
permutations of the spin-1/2 states, written here as tensor product of
the computational basis states $\ket{0}$ and $\ket{1}$ of each
spin-1/2. The Dicke states are in 1-1
correspondence with the $\ket{j,m}$ states, with $j=N/2$ and
$m=k-N/2$.


It was shown in \cite{GBBBM} that a tight frame of matrices
$S_{i_1\ldots i_N}$ can be obtained by projecting
$\boldsymbol{\sigma}_{i_1i_2\ldots i_N}\equiv\sigma_{i_1}\otimes\sigma_{i_2}\ldots\otimes\sigma_{i_N}$
into $\mathcal{H}_S$. More precisely, the
$S_{i_1i_2\ldots i_N}$ are the $(N+1)$-dimensional
blocks spanned by the $\ket{D_N^{(k)}}$ ($k=0,1,\ldots,N$) of the matrix
$\mathcal{P}_S\,\boldsymbol{\sigma}_{i_1i_2\ldots i_N}\,\mathcal{P}_S^\dagger$,
i.e.~in terms of matrix elements
\begin{equation}
\label{Squbit}
\langle
D_N^{(k)}|S_{i_1i_2\ldots i_N}|D_N^{(l)}\rangle=\langle
D_N^{(k)}|\boldsymbol{\sigma}_{i_1i_2\ldots i_N}|D_N^{(l)}\rangle\,.
\end{equation}
By definition, there are $4^N$ matrices
$S_{i_1i_2\ldots i_N}$. However, since they are invariant under
permuation of indices, many of them coincide. $S_{0\ldots 0}$ is the
identity matrix acting on $\mathcal{H}_S$.
Due to the tight-frame property, one can expand any density operator
of a spin-$j$ as
\begin{equation}
\rho = \sum_{i_1,...,i_N=0}^n \frac{1}{2^{N}}\,a_{i_1i_2\ldots i_{N}}S_{i_1i_2\ldots i_{N}},
\label{canonrhoj}
\end{equation}
with real and permutationally invariant coefficients
\begin{equation}
\label{defcoor}
a_{i_1i_2\ldots i_{N}}=\tr(\rho\, S_{i_1i_2\ldots i_{N}})\,.
\end{equation}
Therefore, each density matrix $\rho $ corresponds to a $4$-dimensional tensor
$\mathcal{A}_{N,4}=\left( a_{i _{1}i _{2}\ldots i _{N}}\right) $. Note that there are other
ways than \eqref{defcoor} to choose the $a_{i_1,...,i_N}$ as the
$S_{i_1\ldots i_N}$ form an overcomplete basis. \\

The representing tensor of a coherent state is particularly simple: Since any
spin-$j$ coherent state $\ket{\alpha}$ can be obtained by acting with
$U=e^{-\imath \gamma\bn\cdot \bJ}$ on
$\ket{j,j}=\ket{\frac{1}{2},\frac{1}{2}}^{\otimes N}$,
a spin-$j$
coherent
state is simply a tensor product of spin-1/2 coherent states,
$\ket{\alpha}_j=\ket{\alpha}_{1/2}\otimes\ldots\otimes\ket{\alpha}_{1/2}
$, where we have added a subscript indicating the total spin-quantum
number.  Since it is a
symmetric state ($\mathcal{P}_S\ket{\alpha}=\ket{\alpha}$) we have
\begin{align} \label{S_sigma}
\langle \alpha |S_{i _{1}i _{2}\ldots i _{N}}|\alpha \rangle &=\langle
\alpha |\mathcal{P}_S\boldsymbol{\sigma}_{i _{1}i _{2}\ldots i
  _{N}}\mathcal{P}_S^\dagger|\alpha \rangle =\bra{\alpha}\otimes\ldots\otimes\bra{\alpha}
\sigma_{i_{1}}\otimes\sigma_{i _{2}}\ldots\sigma_{i_{N}}\ket{\alpha}\otimes\ldots\otimes|\alpha \rangle\\
=v_{i _{1}}v_{i
_{2}}\ldots v_{i _{N}}.
\end{align}
As a consequence, $\rho=\ketbra{\alpha}$ has the tensor
representation $a_{i_1\ldots i_N}=v_{i_1}\ldots v_{i_N}$,
i.e.~the representing tensor $\mathcal{A}$ of $\rho=\ketbra{\alpha}$ is a
rank-1 tensor with $v_0=1$ and $||\bv||=1$.

For an arbitrary density matrix $\rho$, the tensor $\mathcal{A}_{N,4}$
enjoys useful properties.  Firstly, the $a_{i_1i_2\ldots i_{N}}$ in Eq.~\eqref{defcoor} are such that
\begin{equation}
a_{00 i_3 \ldots i_N}=\sum_{i=1}^{3}a_{i i\, i_3 \ldots i_N}.  \label{eq:nov14}
\end{equation}
To see this, let $\ket{\alpha}$ be a coherent state.  Since its
representing tensor is $a_{i_1\ldots i_N}=v_{i_1}\ldots v_{i_N}$, and $\bv^2=v_0^2=1$, we have
\begin{equation}
v_{0}v_{0}v_{i _{3}}\ldots v_{i _{N}}=\sum_{a=1}^{3}v_{a}v_{a}v_{i
_{3}}\ldots v_{i _{N}}\,,
\end{equation}%
which is Eq.~\eqref{eq:nov14} for coherent states.
Due to the linearity of the decomposition \eqref{eq:P} of $\rho$ in terms of
coherent states, Eq.~(\ref{eq:nov14}) for arbitrary states follows.

Secondly,  by Eqs. (\ref{eq:P}), (\ref{defcoor}) and (\ref{S_sigma}), we have
\begin{eqnarray}
a_{00\ldots0}
&=& \tr(\rho\, S_{00\ldots 0})\nonumber
\\
&=& \tr\left( \int_{S^2} d\alpha P(\alpha)\ketbra{\alpha} S_{00\ldots 0}\right)\nonumber
\\
&=&  \int_{S^2} d\alpha P(\alpha)\langle\alpha|S_{00\ldots 0}|\alpha\rangle\nonumber
\\
&=&\int_{S^2} d\alpha P(\alpha)\langle\alpha|\boldsymbol{\sigma}_{00\ldots 0}|\alpha\rangle\nonumber
\\
&=& \int_{S^2} d\alpha P(\alpha),
\end{eqnarray}
so that $a_{00\ldots0}=1$ if the state is normalized.
Finally, as
shown in \cite{GBBBM}, the
$a_{i_1i_2\ldots i_{N}}$ are unique if they  are restricted to real numbers, invariant under
permutation of the indices, and verifying the condition  Eq.~(\ref{eq:nov14}).
There is therefore a mapping from  { the density matrices $\rho $ of a
spin-$j$ state to  $4$-dimensional real symmetric normalized tensors of
order $N=2j$, $\mathcal{A}_{N,4}=\left( a_{i _{1}i _{2}\ldots i _{N}}\right) \in
S_{N,4}$. We call this tensor the ``representing tensor'' of the state
$\rho$.}  





Hence, by Eq.~\eqref{eq:Pw}, a spin-$j$ state is classical  if and only if there are { positive weights $w_k > 0$ for $k=1, \ldots, r$}, and vectors $\vv^{(k)} = \left(1, v^{(k)}_1, v^{(k)}_2, v^{(k)}_3\right)^\top \in \Re^{ 4}$, satisfying
\begin{equation} \label{e11}
\left(v^{(k)}_1\right)^2 + \left(v^{(k)}_2\right)^2 + \left(v^{(k)}_3\right)^2=1,
\end{equation}
for $k = 1, \ldots, r$, such that the representing tensor  $\A =
(a_{i_1\ldots i_N}) \in S_{N, 4}$ of that spin-$j$ state satisfies
\begin{equation} \label{e12}
\A = \sum_{k=1}^r w_k \left(\vv^{(k)}\right)^{\otimes N},
\end{equation}
i.e., $\A$ is a regularly decomposable tensor.

\bigskip

Based upon the above discussions and Theorem \ref{t1}, we have the following theorem.

\begin{theorem} \label{t2}
{ The tensor   $\A = (a_{i_1\ldots i_N}) \in S_{N, 4}$ representing   a spin-$j$ state (with $N = 2j$) is a regular symmetric tensor.   A
  spin-$j$ state is classical if and only if its representing tensor
  is a regularly decomposable tensor.}
\end{theorem}

Thus, the physical problem  { of determining whether} a spin-$j$
state is classical or not is equivalent to a mathematical problem to
determine {whether its representing tensor
is a regularly} 
decomposable tensor or not.

\section{Properties of Completely Decomposable and Regularly Decomposable Tensors}

There {is already substantial literature
on PSD tensors and SOS tensors, including}
\cite{CLQ, CQ, LWZZL,
  LQX, Qi, Qi15, QS, QXX, SQ, WLQX, ZQZ}.   There are only two papers
on completely decomposable tensors \cite{LQX, WLQX}.    Regularly
decomposable tensors are introduced in this paper.
By the discussion in the last section, we see that regularly
decomposable tensors play a significant role { for the} classicality of spin
states.  Thus, in this section, we discuss properties of completely
decomposable tensors and regularly decomposable tensors.

\subsection{Invariance of complete decomposability and regular decomposability}
Any measure of entanglement should be invariant under local unitary
transformations { (see e.g.~\cite{Plenio05}).}
 Hence, also the set of
fully separable
states must be invariant under local unitary transformations.
Correspondingly, the classicality of a spin-$j$ state should be
invariant under rotations of the coordinate system. For a physical
system in three spatial dimensions, such a rotation is represented by the
3$\times$3 orthogonal transformation matrix $R(\bn,\gamma)$
introduced above that acts on a vector of spatial coordinates
$x_1,x_2,x_3$. The corresponding transformation of a covariant tensor
(i.e.~a tensor that transforms as the coordinates) of
dimension $4$ and order $m$ is given by its { inner product with  ${\mathcal R}^{\otimes m}$, where ${\mathcal R}$
is defined by Eq.~\eqref{Rs}}.
More generally, we expect the regular decomposability
of a tensor to be a property
invariant under  orthogonal transformations
described by an $(n+1) \times (n+1)$ matrix
$${\mathcal R} = \left({\begin{array} {cc} 1 & \0^\top \\ \0 & R \end{array}}\right),$$
where $\0$ is the zero vector in $\Re^n$, and $R$ is now an $n \times
n$ orthogonal matrix.   Then
$$R\left({\begin{array} {c} x_1 \\ x_2 \\ . \\ . \\ . \\ x_{n-1} \\ x_n \end{array}}\right) = \left({\begin{array} {c} y_1 \\ y_2 \\ . \\ . \\ . \\ y_{n-1} \\ y_n \end{array}}\right)$$
and
$${\mathcal R}\left({\begin{array} {c} x_0 \\ x_1 \\ x_2 \\ . \\ . \\ . \\ x_{n-1} \\ x_n \end{array}}\right) = \left({\begin{array} {c} y_0 \\ y_1 \\ y_2 \\ . \\ . \\ . \\ y_{n-1} \\ y_n \end{array}}\right)$$
with $x_0 = y_0$.   We call such an orthogonal matrix a {\bf
  normalized orthogonal matrix}.   Denote ${\mathcal R} = (r_{li})$.
As in \cite{Qi}, for any symmetric tensor $\A = (a_{i_1\ldots i_m})
\in S_{m, n+1}$, let $\B = (b_{l_1\ldots l_m}) \equiv {\mathcal R}^m\A \in S_{m, n+1}$ be defined by
$$b_{l_1\ldots l_m} = \sum_{i_1, \ldots, i_m = 0}^n a_{i_1\ldots i_m}r_{l_1i_1}\ldots r_{l_mi_m}$$
for $l_1, \ldots, l_m = 0, \ldots, n$.   By \cite{Qi}, $\A$ and $\B$
have the same E-eigenvalues and Z-eigenvalues.   In particular, when
$m$ is even, $\A$ is PSD if and only if $\B$ is PSD.    By \cite{LQY},
when $m$ is even, $\A$ is SOS if and only if $\B$ is SOS.  This shows
that the PSD property and the SOS property can represent physical properties, as
they are invariant under orthogonal transformation.

\begin{theorem} \label{t4.1}
Let ${\mathcal R}$ be a normalized orthogonal matrix, $\A, \B \in
S_{m, n+1}, \B = {\mathcal R}^m\A$.   Then $\A$ is completely decomposable if and only if $\B$ is completely decomposable, and $\A$ is regularly decomposable if and only if $\B$ is regularly decomposable.
\end{theorem}

 \noindent {\bf Proof.}  Suppose that $\A = (a_{i_1\ldots i_m}) \in S_{m, n+1}$ is completely decomposable, $\B = (b_{k_1\ldots k_m}) \in S_{m, n+1}$, $\B = {\mathcal R}^m \A$, where ${\mathcal R} = (r_{li})$ is an $(n+1) \times (n+1)$ orthogonal matrix.  Then there are vectors $\uu^{(1)}, \ldots, \uu^{(r)} \in \Re^{n+1}$, where
 $\uu^{(k)} = (u^{(k)}_0, \ldots, u^{(k)}_n)^\top$ for $k=1, \ldots, r$, such that
$$\A = \sum_{k=1}^r \left(\uu^{(k)}\right)^{\otimes m},$$
i.e., for $i_1, \ldots, i_m = 0, \ldots, n$,
$$a_{i_1\ldots i_m} = \sum_{k=1}^r u^{(k)}_{i_1} \ldots u^{(k)}_{i_m}.$$
Then, for $l_1, \ldots, l_m = 0, \ldots, n$, we have
\begin{eqnarray*}
b_{l_1\ldots l_m} &=&
\sum_{i_1, \ldots, i_m = 0}^n a_{i_1\ldots i_m}r_{l_1i_1}\ldots r_{l_mi_m}
\\
&=& \sum_{k=1}^r \sum_{i_1, \ldots, i_m = 0}^n u^{(k)}_{i_1} \ldots u^{(k)}_{i_m}r_{l_1i_1}\ldots r_{l_mi_m}
\\
&= &  \sum_{k=1}^r v^{(k)}_{l_1} \ldots v^{(k)}_{l_m},
\end{eqnarray*}
where for $k= 1, \ldots, r, l = 0, \ldots, n$,
$$v^{(k)}_l = \sum_{i=0}^n r_{li}u^{(k)}_i.$$
This implies that
$$\B = \sum_{k=1}^r \left(\vv^{(k)}\right)^{\otimes m},$$
where $\vv^{(k)} = (v^{(k)}_0, \ldots, v^{(k)}_n)^\top$ for $k = 1, \ldots r$.  This implies that $\B$ is completely decomposable.  By  \cite{Qi}, if
$\B = {\mathcal R}^m \A$, then $\A = ({\mathcal R}^\top)^m \B$.   Thus, if $\B$ is completely decomposable, then $\A$ is also completely decomposable.

Assume that $m$ is even, $\A$ is regularly decomposable and ${\mathcal R}$ is a normalized orthogonal matrix.   Then, we may assume that in the above discussion, vectors $\uu^{(1)}, \ldots, \uu^{(r)}$ are regular.
Since $\vv^{(k)} = {\mathcal R}\uu^{(k)}$ for $k = 1, \ldots, r$, and
${\mathcal R}$ is a normalized orthogonal matrix, we may conclude that
$\vv^{(1)}, \ldots, \vv^{(r)}$ are also regular.
This implies that $\B$ is also regularly decomposable.   By  \cite{Qi}, if
$\B = {\mathcal R}^m \A$, then $\A = ({\mathcal R}^\top)^m \B$.   Thus, if $\B$ is regularly decomposable, then $\A$ is also regularly decomposable.

Now assume that $m$ is odd, $\A$ is regularly decomposable and ${\mathcal R}$ is a normalized orthogonal matrix.  Then $\B_0$ is also regularly decomposable.    As $\A_i$ for $i = 1, \ldots, n$, are induced from the regular decomposition of $\A_0$, we may see that $\B_i$ for $i = 1, \ldots, n$, are induced from the regular decomposition of $\B_0$.   This implies that $\B$ is also regularly decomposable.   Similarly, if $\B$ is regularly decomposable, then $\A$ is also regularly decomposable.
 \qed
 
 The proof of this theorem can be simplified by applying Theorem 2.2 of \cite{LQY}, Theorem 2.4 and the definition of the normalized orthogonal matrices in this paper.

 These show that complete decomposability and regular decomposability
 are invariant under normalized orthogonal transformation.

 \subsection{Hadamard Products}

For any two tensors $\A = (a_{i_1\cdots i_m})$, $\B = (b_{i_1\cdots i_m}) \in T_{m,
n+1}$, their {\bf Hadamard product}, denoted as $\A \circ \B$, \index{Hadamard product} is
defined by
\begin{equation}\label{hadamardproduct}\A \circ \B = (a_{i_1\cdots i_m}b_{i_1\cdots i_m}) \in T_{m, n+1}.\end{equation}

In matrix theory, the Hadamard product of two PSD symmetric matrices is also a PSD symmetric matrix.   This is no longer true for tensors.   In \cite{Qi15}, an example was given that  the Hadamard product of two PSD Hankel tensors may not be PSD.  Hankel tensors are symmetric tensors.   Thus,
the Hadamard product of two PSD symmetric tensors may not be PSD.   In \cite{LQY}, an example was given that the Hadamard product of two SOS tensors may not be an SOS tensor.     However, we have the following proposition:

\begin{proposition} \label{p4.1}   Suppose that $\A = (a_{i_1\cdots i_m})$, $\B = (b_{i_1\cdots i_m}) \in S_{m, n+1}$ are completely decomposable tensors.   Then their Hadamard product $\A \circ \B$ is also a completely decomposable tensor.
\end{proposition}

 \noindent {\bf Proof.}  Suppose that $\A$ and $\B$ are completely decomposable.   Then there are vectors $\uu^{(1)}, \ldots, \uu^{(r)}, \vv^{(1)}, \ldots, \vv^{(p)} \in \Re^{n+1}$, such that
$$\A = \sum_{k=1}^r \left(\uu^{(k)}\right)^{\otimes m}$$
and
$$\B = \sum_{l=1}^p \left(\vv^{(l)}\right)^{\otimes m}.$$
Then is easy to see that
$$\A \circ \B = \sum_{k=1}^r \sum_{l=1}^p \left(\uu^{(k)} \circ \vv^{(l)} \right)^{\otimes m},$$
i.e., $\A \circ \B$ is completely decomposable.
 \qed

 This property is no longer true for regularly decomposable tensors.   In this sense, completely decomposable tensors are similar to completely positive tensors studied in \cite{QXX}: the Hadamard product of two completely positive tensors is still a completely positive tensor.

 \subsection{Duality between the PSD Tensor Cone and the Completely Decomposable Tensor Cone}

 Denote the set of all completely decomposable tensors in $S_{m, n+1}$
 by $CD_{m, n+1}$, the set of all regularly decomposable tensors in
 $S_{m, n+1}$ by {$RD_{m, n+1}$.}
   Let $m$ be even, denote the set of all PSD tensors in $S_{m, n+1}$
   by $PSD_{m, n+1}$, the set of all SOS tensors in $S_{m, n+1}$ by
   $SOS_{m, n+1}$.   Then $CD_{m, n+1}$, $RD_{m, n+1}$, $PSD_{m, n+1}$, and
   $SOS_{m, n+1}$ are cones. 

 Let $C$ be a cone in $S_{m, n+1}$.   Then its dual cone $C^*$ is defined by
 $$C^* : = \{ \A \in S_{m, n+1} : \A \bullet \B \ge 0, \ {\rm for\ all}\ \B \in C \}.$$
 The dual cone $C^*$ is a closed convex cone.   The dual cone of $C^*$
 is the closure of the convex hull of $C$.   If $C$ is closed and
 convex, then $C$ and $C^*$ are dual cones to each other.  Let $\A = (a_{i_1\ldots i_m}) \in S_{m, n+1}$ and $\Re^{n+1}_+$ be the nonnegative orthant of $\Re^{n+1}$.  If for any $\x \in \Re^{n+1}_+$, we have $\A \bullet \x^{\otimes m} \ge
0$, then we say that $\A$ is a {\bf copositive}
tensor.  Copositive tensors have also applications in physics \cite{Ka}.}      By
 \cite{QXX}, the completely positive tensor cone and copositive  tensor cone are dual cones to each other.

 By \cite{QY} and the definition of completely decomposable tensors, we have the following proposition.   A part of this proposition is covered by Proposition 4.2 of \cite{LQY}.

 \begin{proposition} \label{p4.2}
 Suppose that $m$ is even.   Then $PSD_{m, n+1}$ and $CD_{m, n+1}$ are dual cones to each other.  Thus, both are closed convex cones.
 \end{proposition}

 \subsection{Closedness and Convexity of the Regularly Decomposable Tensor Cone}

 In the last subsection, we already knew that if $m$ is even, then $PSD_{m, n+1}$ and $CD_{m, n+1}$ are closed convex cones.   By \cite{LQX}, if $m$ is odd, $CD_{m, n+1}$ is the linear space $S_{m, n+1}$.    By \cite{LQY}, $SOS_{m, n+1}$ is also a closed convex cone.  We now discuss closedness and convexity of $RD_{m, n+1}$.

 \begin{proposition} \label{p4.3}
 $RD_{m, n+1}$ is a closed convex cone.
 \end{proposition}

 \noindent {\bf Proof.}   Suppose that $\{ \A^{(l)} : l = 1, 2, \ldots, \}$ is a sequence of regularly decomposable tensors in $RD_{m, n+1}$ such that
 $$\A = \lim_{l \to \infty} \A^{(l)}.$$
 By Theorem \ref{t1}, we may assume that
 $$\A^{(l)} = \sum_{k=1}^{r_l} \alpha_{k, l}\left(\vv^{(k, l)}\right)^{\otimes m},$$
where $\alpha_{k, l} \ge 0$, $\vv^{(k, l)} = (1, v^{(k, l)}_1, \ldots, v^{(k, l)}_1)^\top$,
$$\left(v^{(k, l)}_1\right)^2 + \ldots + \left(v^{(k, l)}_n\right)^2 = 1,$$
for $k = 1, \ldots, r_l$, for $l = 1, 2, \ldots$.   By the Carath\'{e}odory theorem, we may assume that
$$r_l \le R \equiv \left({n+m+2 \atop m}\right) + 1.$$
Thus, by taking a subsequence if necessary, without loss of generality, there is a $r \le R$ such that $r_l = r$ for $l = 1, 2, \ldots$.  Then, we may conclude that there are $\alpha_k \ge 0$, $\vv^{(k)} = (1, v^{(k)}_1, \ldots, v^{(k)}_1)^\top$,
$$\left(v^{(k)}_1\right)^2 + \ldots + \left(v^{(k)}_n\right)^2 = 1,$$
for $k = 1, \ldots, r$.   Thus, by Theorem \ref{t1}, $\A$ is a regularly decomposable tensor.   This shows that
$RD_{m, n+1}$ is a closed cone.   Following directly from Theorem \ref{t1},  we see that  $RD_{m, n+1}$ is also a convex cone.
 \qed

\section{Concluding Remarks}

In this paper, we have {introduced the concept of regularly
  decomposable tensors.  We have shown}
that a spin state is classical if and only if
its representing tensor is a regularly decomposable tensor.   Thus,
the problem for determining whether a spin state is classical or not is
mathematically equivalent to the problem of determining whether a given
{ regular} symmetric tensor is a regularly decomposable tensor or not.

How can we construct an algorithm for determining a given regular
symmetric tensor is a regularly decomposable tensor or not?   We see
that the properties of completely decomposable tensors and regularly
decomposable tensors in some extent are similar to {those of }
completely positive tensors \cite{LQ, QXX}.
Recently, an algorithm for determining { whether}  a given
symmetric nonnegative
tensor is completely positive
or not was proposed \cite{FZ}.   Perhaps we may learn from that algorithm
{ how} to construct an algorithm determining whether a given
regular symmetric
 tensor is regularly decomposable or not.

  \end{document}